\documentclass[useAMS,usenatbib,letterpaper]{mn2e}
\usepackage[totalwidth=480pt,totalheight=680pt,layoutvoffset=0.5cm]{geometry}
\usepackage{rotating}
\def\lesssim{\mathrel{\hbox{\rlap{\hbox{\lower3pt\hbox{$\sim$}}}\hbox{\raise2pt\hbox{$<$}}}}}
\def\gtrsim{\mathrel{\hbox{\rlap{\hbox{\lower3pt\hbox{$\sim$}}}\hbox{\raise2pt\hbox{$>$}}}}}

\title[Molecular gas in LBAs]{Molecular gas properties of UV-luminous star-forming galaxies at low redshift}
\author[T. S. Gon\c{c}alves et al.]{Thiago S. Gon\c{c}alves$^{1}$, Antara Basu-Zych$^{2}$, Roderik A. Overzier$^{3}$, Laura P\'erez${^{4,\dagger}}$, 
\newauthor D. Christopher Martin$^{5}$\\
$^{1}$Observat\'orio do Valongo, Universidade Federal do Rio de Janeiro, Ladeira Pedro Antonio 43, Sa\'ude, Rio de Janeiro-RJ, CEP 20080-090, Brazil\\
$^{2}$NASA Goddard Space Flight Center, Laboratory for X-ray Astrophysics, Greenbelt, MD 20771, USA\\
$^{3}$Observat\'orio Nacional, Rua Jos\'e Cristino 77, S\~ao Crist\'ov\~ao, Rio de Janeiro-RJ, CEP 20921-400, Brazil \\
$^{4}$Jansky Fellow, National Radio Astronomy Observatory, Socorro, NM 87801, USA\\
$^{5}$California Institute of Technology, MC 278-17, 1200 East California Boulevard, Pasadena, CA 91125, USA}

\begin{document}

\date{\today}

\pagerange{\pageref{firstpage}--\pageref{lastpage}} \pubyear{2013}

\maketitle

\label{firstpage}

\begin{abstract}

Lyman break analogues (LBAs) are a population of star-forming galaxies at low redshift ($z\sim 0.2$) selected in the ultraviolet (UV). These objects present higher star formation rates and lower dust extinction than other galaxies with similar masses and luminosities in the local universe. In this work we present results from a survey with the Combined Array for Research in Millimetre-wave Astronomy (CARMA) to detect CO(1-0) emission in LBAs, in order to analyse the properties of the molecular gas in these galaxies. Our results show that LBAs follow the same Schmidt-Kennicutt law as local galaxies. On the other hand, they have higher gas fractions (up to 66\%) and faster gas depletion time-scales (below 1 Gyr). These characteristics render these objects more akin to high-redshift star-forming galaxies. We conclude that LBAs are a great nearby laboratory for studying the cold interstellar medium in low-metallicity, UV-luminous compact star-forming galaxies.

\end{abstract}


\section{Introduction}\label{sec:carma_intro}

\let\thefootnote\relax\footnotetext{$\dagger$ Jansky Fellow}
Molecular gas is one of the most fundamental ingredients in the formation of stars in galaxies. Stars form from the collapse of cold hydrogen gas; therefore H$_2$ can be seen as the primary fuel reservoir for star formation in a system \citep{Leroy2008,Schruba2011,Kennicutt2012}. Nevertheless, although it is so abundant in star-forming galaxies (SFGs) -- sometimes being the dominant component of the baryonic mass \citep{Erb2006,Tacconi2010,Goncalves2011} -- we cannot directly observe it. Since it has no permanent dipole, no rotational millimetre lines are observable, and the direct detection of H$_2$ is very difficult \citep[with the exception of very hot gas outside photodissociation regions; see][]{Zakamska2010}. Therefore, we rely on emission from rotational transitions in carbon monoxide (CO, the most abundant dipolar molecule in molecular clouds) to study the total molecular gas reservoir. In order to determine the molecular gas masses, we then assume that 

\begin{equation}
M({\rm H}_2) = \alpha_{\rm CO}L'_{\rm CO}, \label{eq:alpha_co}
\end{equation}
where gas masses ($M({\rm H}_2)$) are measured in $M_\odot$ and CO luminosities (${\rm L}'_{\rm CO}$) are measured in K km s$^{-1}$ pc$^2$. From dynamical mass measurements, the $\alpha_{\rm CO}$ factor has been determined to be 4.6 in the Milky Way \citep{Solomon1991}. However, using the same factor for ultra-luminous infrared galaxies (ULIRGs) yields gas masses larger than dynamical masses. From kinematic models, \citet{Downes1998} inferred a typical value of  $\alpha_{\rm CO}=0.8$ for these galaxies. The picture is still unclear for high-redshift galaxies; while\citet{Narayanan2012} argue that elevated gas temperatures and velocity dispersion will increase CO luminosities, the opposite is true for galaxies with low metallicities, since the CO molecule in giant clouds is destroyed more easily by the ambient ultraviolet (UV) radiation \citep{Leroy2011, Genzel2012, Bolatto2013}. The combined effect for metal-poor, turbulent SFGs has not yet been studied in detail.

Despite inherent difficulties in determining total masses from carbon monoxide alone, CO observations have been a key instrument to our understanding of the molecular gas distribution of our local universe. It was established early on that a correlation exists between star formation and gas density in the Milky Way \citep{Schmidt1959}. Later work by \citet{Kennicutt1998} based on a combination of CO, HI and far-infrared (FIR) observations determined that the correlation extends to gas and star-formation surface brightness for SFGs, given by

\begin{equation}
\Sigma_{SFR} = A\Sigma_{\rm gas}^N. \label{eq:sklaw}
\end{equation}
This relation is known as the Schmidt-Kennicutt (S-K) law, with constants $A=2.5\times 10^{-4}$  and $N=1.4$ empirically determined by \citet{Kennicutt1998}, for $\Sigma_{\rm gas}$ in units of $M_\odot$ pc$^{-2}$ and $\Sigma_{\rm SFR}$ in units of $M_\odot$ yr$^{-1}$ kpc$^{-2}$.

More recently, \citet{Bigiel2008} analysed data from nearby SFGs to resolve the S-K relation down to scales of 750 pc. These authors found that the relation in the nearby universe can be divided into two: one for atomic gas alone, which saturates at approximately 10 M$_\odot$ pc$^{-2}$; and a linear relation ($N=1$) for molecular gas, mostly above that saturation level. The linearity has an interesting consequence: since the S-K law relates star formation rates (SFR) and the gas reservoir, an $N=1$ index means a constant gas depletion time-scale for molecular gas in observed SFGs, $t_{\rm depletion}=\Sigma_{{\rm H}_2}/\Sigma_{\rm SFR}$. This value was measured to be approximately 2 Gyr, with $1-\sigma$ scatter of around 0.24 dex, or 75\% \citep{Bigiel2008}.

In an attempt to offer a theoretical scenario for the formation of stars in giant molecular clouds, \citet{Krumholz2009} proposed a simple model listing three main factors that determined the gas depletion time-scales: gas fractions, internal star-formation feedback and turbulence. At first the fraction of gas available for star formation is dependent on self-shielding from the interstellar radiation field -- it is this factor that sets the threshold of conversion from atomic to molecular hydrogen. Once molecular clouds are formed, feedback determines their properties, since the internal pressure is higher at this stage than the average interstellar medium (ISM) pressure. Ongoing star formation in these objects is regulated by turbulence to a universal rate of 1\% of the mass per free-fall time. These can be combined into an analytic formalism capable of predicting the linear correlation between gas surface densities and constant depletion time-scales.

Understanding the detailed process through which molecular gas translates into stars becomes even more complicated for extreme objects, as has been shown for ULIRGs \citep[$L_{\rm bol}>10^{12}$ $L_\odot$;][]{Sanders1996} both at low and high redshifts. The observed gas density in these galaxies is well above the limit of $\sim 100$ M$_\odot$ pc$^{-2}$ mentioned in the work of \citet{Krumholz2009}, which means the assumption of the properties in giant molecular clouds being determined by internal dynamics is no longer valid, and the end result is likely a distinct scenario for conversion of gas into stars.

In an effort to understand the conversion of gas into stars in these environments, \citet{Bouche2007} and \citet{Tacconi2008} have measured the surface density of molecular gas in submillimetre-selected high-redshift ULIRGs, and have determined that the exponent in the S-K law seems to be higher for these objects, i.e. the star formation surface density is higher than expected for a given value of gas surface density. The same is true for intermediate-redshift ULIRGs: \cite{Combes2010} have measured CO luminosities of 30 galaxies between $0.2<z<0.6$ and have found star formation efficiencies three times higher than those found in the local universe.

\citet{Daddi2010} have done the same exercise for a number of SFGs at $z\sim 2$ selected in the optical and near-infrared (the $BzK$ sample), which present more regular structures, with signs of a rotating gas disc \citep[see also][]{Tacconi2013}. Although these galaxies present higher SFR and gas densities than spirals in the local universe, they appear to follow the same gas--star formation relation as their low-redshift counterparts. In a later paper, \citet{Daddi2010a} have argued for the existence of a `bimodal' S-K law, with distinct normalizations for disc galaxies \citep[now commonly referred to as `main-sequence SFGs'; e. g.,][]{Noeske2007,Rodighiero2011} and dusty, ultraluminous starbursts. Interestingly, this bimodality ceases to exist once the dynamical times in such galaxies is taken into account, i.e., the universal S-K law can be written as

\begin{equation}
\Sigma_{SFR} = \alpha\left(\frac{\Sigma_{gas}}{t_{dyn}}\right)^{\nu},\label{eq:sklaw_tdyn}
\end{equation}
where $\alpha$ and $\nu$ are different proportionality constants, and $t_{dyn}$ is the dynamical time-scale of the galaxy.

Nevertheless, these studies are still biased towards more massive galaxies, simply because smaller objects are faint and difficult to observe. These massive objects, in turn, tend to be dynamically colder, with more significant rotational support and a more metal-rich ISM than less massive ones \citep{Maiolino2008,Goncalves2010}. Moreover, they probably do not represent the `typical' SFG at high-$z$, since the mass function of galaxies is steep at these epochs \citep{Reddy2009}.

Until recently, only two Lyman break galaxies (LBGs) had been observed, both of which are strongly lensed and therefore highly magnified \citep{Baker2004,Coppin2007,Riechers2010}. Later, \citet{Magdis2012a} have detected one unlensed LBG; this object, however, is very massive, with log $M_*/M_\odot \sim 10^{11}$. In order to understand how the star formation process operates on all scales, on the other hand, it is important to study SFGs at all masses. We conclude that our understanding of the conversion of gas into stars in such conditions could benefit greatly from the detailed study of the cold molecular gas in low-redshift galaxies with properties similar to those commonly found around $z\sim 2$, allowing us to target multiple objects with less extreme CO luminosities which are likely more representative of the typical SFGs in these epochs.

Lyman break analogues (LBAs) offer an elegant solution to this problem. These objects, present at $z\sim 0.2$, show a number of properties that render them more akin to high-redshift UV-selected SFGs than normal spirals in the local universe (see Section \ref{sec:lba_sample}). Therefore, we are able to study the cold ISM in these galaxies in order to infer how the conversion of the molecular gas reservoir into stars happens under such conditions.



In this work, we present results from a molecular gas survey of five LBAs, observed with the CARMA interferometer. We target the lowest CO transition, CO(1-0), to most accurately trace the cold molecular gas in these galaxies. This paper is divided as follows: in Section \ref{sec:carma_data}, we present details on sample selection, observations and data reduction; in Section \ref{sec:carma_results}, we present the results from our survey, comparing properties such as gas masses and gas fractions with those found at high redshift; and in Section \ref{sec:carma_discussion} we discuss our results, with caveats (including the implications of different $\alpha_{\rm CO}$ factors at low metallicities) and future perspectives on our work. We summarize our results in Section \ref{sec:carma_summary}.

\section{Observations and data reduction}\label{sec:carma_data}

\subsection{The sample of LBAs}\label{sec:lba_sample}

\citet{Heckman2005} and \citet{Hoopes2007} have defined a sample of ultraviolet-luminous galaxies (UVLGs) as those objects with extreme far-ultraviolet (FUV) luminosities, i.e. $L_{\rm FUV}>2\times 10^{10} L_\odot$ at redshifts $z\sim 0.2$ from a combination of the SDSS \citep{Abazajian2009} and {\it GALEX} catalogues \citep{Martin2005}. In order to exclude large spirals and select those galaxies with similar properties to high-redshift SFGs, these authors have further refined their criteria to include only galaxies with high values of surface brightness, $I_{\rm FUV}>10^{9} L_\odot$ kpc$^{-2}$.

Subsequent work has shown that these supercompact UVLGs do indeed resemble high-redshift UV-luminous SFGs, especially the well-studied sample of LBGs \citep[see][and references therein]{Shapley2011}. \citet{Hoopes2007} demonstrated that supercompact UVLGs shared many properties with LBGs: generally speaking, they are more luminous in UV bands, more metal poor, present higher SFR and less dust extinction than typical galaxies with similar masses at low redshift, being instead more similar to their counterparts at $z\sim 2$. This justifies our denomination of the sample as LBAs.

\citet{Overzier2010} showed that the morphologies of LBAs are also distinct. By using quantitative measurements of their morphological structure \citep[such as Gini and M20,][]{Lotz2006}, the authors have shown that these galaxies are irregular, and smaller than galaxies with similar masses at low redshift (with typical half-light radii $R_{50}\sim 1$ kpc). Simulated observations of the sample at $z\gtrsim 2$ showed that LBAs are very similar to LBGs, with many of the asymmetries and giant star-forming clumps regularly seen at high redshift \citep[e.g.,][]{Elmegreen2009}.

Similarly, IFU studies of the ionized gas in LBAs have shown similarities in the kinematic structures between both samples -- with ratios between rotational velocities and dispersion in the gas of $v_{\rm circ}/\sigma \lesssim 1$, much less than typical values above 10 in local spirals -- although there are indications that observational limitations at high redshift, such as the loss in spatial resolution, makes objects appear more symmetric than in reality \citep{Goncalves2010}.

More recently, a study on the infrared properties of LBAs has shown that, while these objects have bolometric luminosities similar to local luminous infrared galaxies (LIRGs), with typical values above L$_{\rm IR}>10^{11}$ L$_\odot$, their dust obscuration, as measured by the infrared excess L$_{\rm IR}$/L$_{\rm FUV}$, are at least an order of magnitude weaker \citep{Overzier2011}. Again, these values are very similar to what is found in LBGs.

Finally, recent work by \citet{Basu-Zych2013} using the {\it Chandra} telescope has shown that the same analogy applies to the X-ray properties of LBAs. These galaxies show elevated X-ray luminosities, with L$_{\rm X}\sim 4$ times the typical value of galaxies with similar SFR in the local universe. The authors argue that this is a result of the low-metallicity of the objects driving the formation of luminous high-mass X-ray binaries. Once more, this is in good agreement with findings for high-redshift SFGs \cite[e.g.][]{Basu-Zych2013a}.	


\subsection{CARMA sample selection}\label{sec:carma_sample}

Our sample was drawn from the same parent sample as that first presented in \citet{Overzier2009}. These objects have been extensively studied with multiple instruments and in a wide variety of wavelengths \citep[e.g.,][]{BasuZych2007,Goncalves2010,Overzier2010,Heckman2011,Alexandroff2012}, and as such will provide useful data in future work comparing other physical properties, such as gas kinematics, with the molecular gas observations presented here.

In an effort to optimize the CO detection of our pilot programme, we further constrained our sample to LBAs with estimated total fluxes above $S_{\rm CO} > 1.0$ Jy km s$^{-1}$. This was calculated by inverting the S-K relation \citep{Kennicutt1998}. If the gas surface density and star formation surface density are related according to Equation \ref{eq:sklaw}, then the total gas in a galaxy will be given by

\begin{equation}
M_{gas} = \frac{\rm SFR}{A}\left(\pi r^2\right)^{(N-1)/N}, \label{eq:gasmass}
\end{equation}
assuming the relation between CO luminosities and total mass is given by Equation \ref{eq:alpha_co}.

In this work we have assumed $\alpha_{\rm CO}=4.6$, to ensure consistency with mass values determined at low and high redshift. This is also consistent with most high-redshift studies of `main-sequence' galaxies \citep{Daddi2010,Tacconi2010,Magdis2012a}. We emphasize that, given their SFR and stellar masses, galaxies in our sample would be classified as starbursts, with typical specific star formation rates sSFR $\equiv$ SFR/M* $\sim 1$ Gyr$^{-1}$, well above other `main-sequence' objects at $z\sim 0.2$ \citep{Elbaz2011}. Nevertheless, they are remarkably distinct from typical dusty starbursts at the same epoch, instead following the same relation between infrared excess and UV slope as other low-redshift SFGs and high-redshift LBGs \citep{Meurer1999,Overzier2011}. Therefore, we consider our galaxies to be more akin to early `main-sequence' objects, justifying our choice of $\alpha_{\rm CO}$. We discuss the implications and possible caveats associated with this selection in Section \ref{section:co_to_h2}.

The conversion between CO luminosities and line flux is

\begin{equation}
L'_{\rm CO} = 3.25\times10^7 \; S_{\rm CO} \Delta\nu \;\nu_{\rm obs}^{-2} \; D_{\rm L}^2 \;(1+ z)^{-3}, \label{eq:sco_lco}
\end{equation}
where $L'_{\rm CO}$ is the CO luminosity in units of K km s$^{-1}$ pc$^2$ and $S_{\rm CO} \Delta\nu$ is the measured flux in Jy km s$^{-1}$. $\nu_{\rm obs}$ is the observed frequency of the CO(1--0) transition and $D_{\rm L}$ is the luminosity distance to the galaxy in Mpc \citep{Solomon2005}.

\begin{figure*}
\begin{center}
\includegraphics[width=0.3\linewidth]{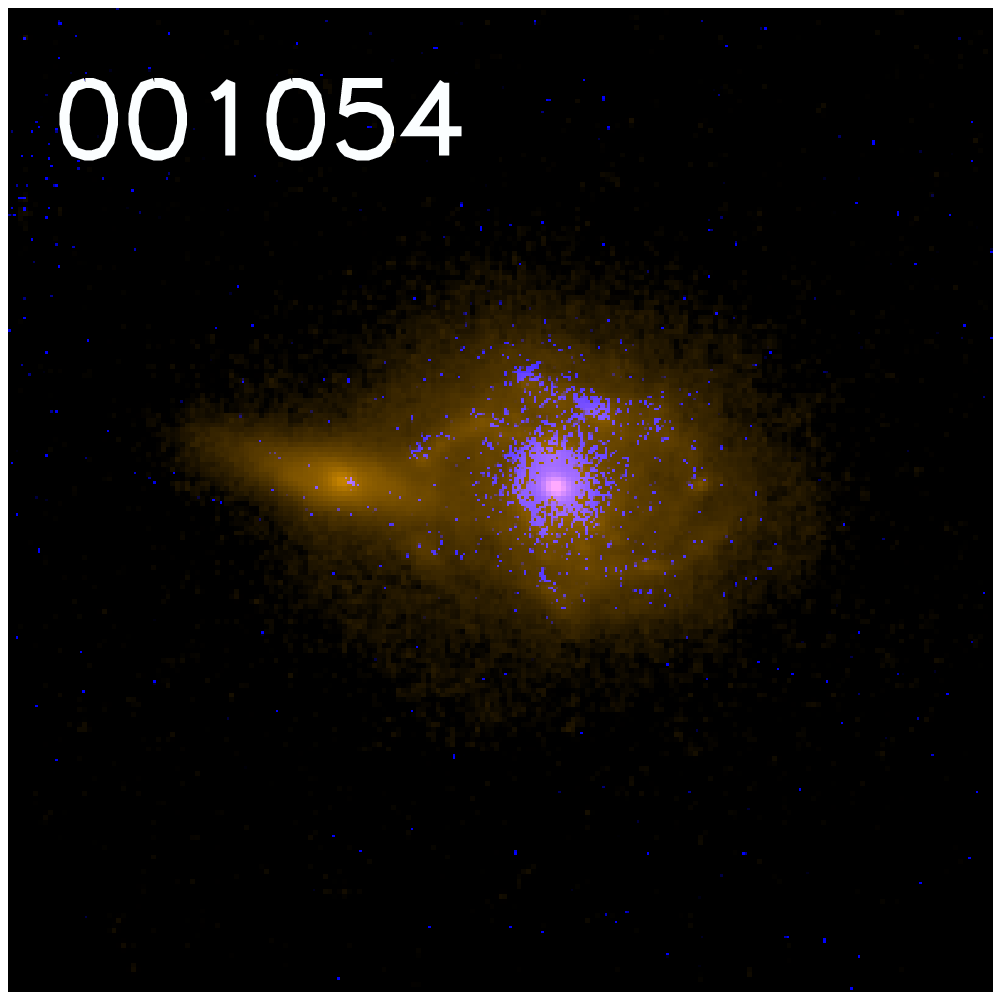}
\includegraphics[width=0.3\linewidth]{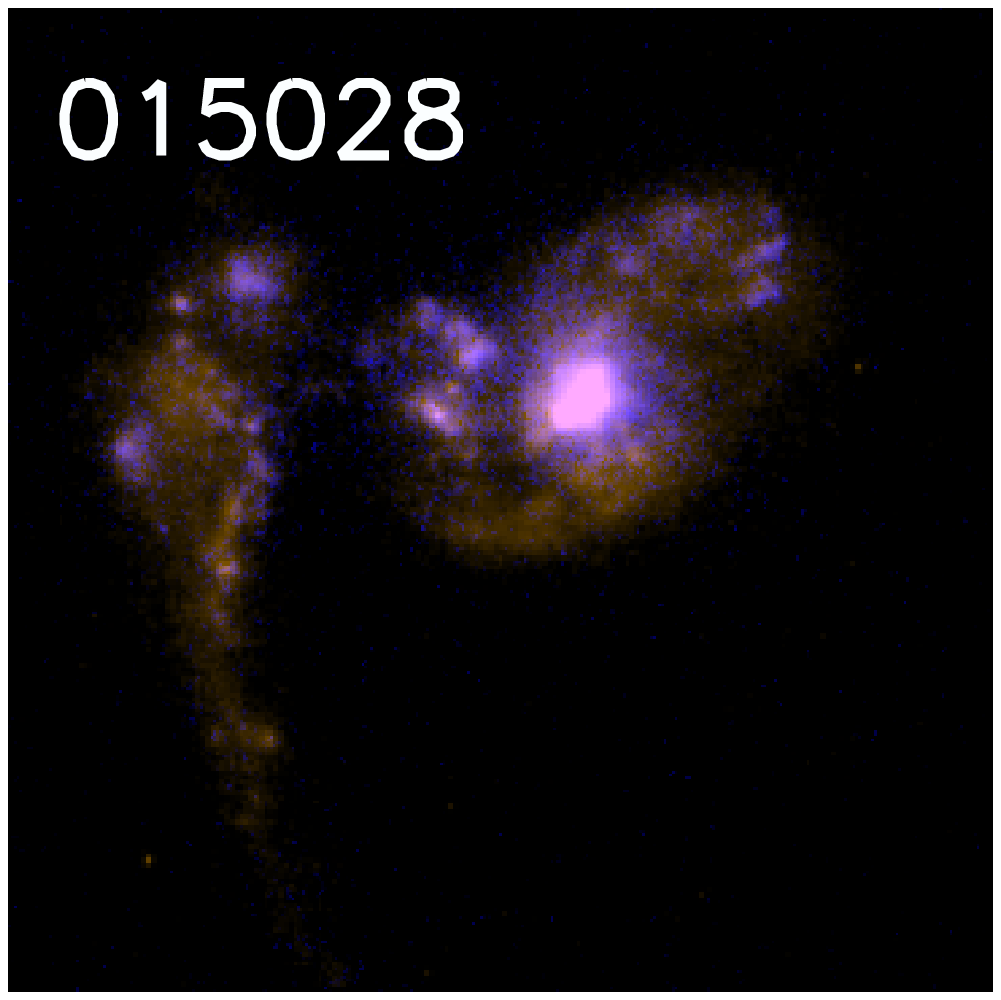}
\includegraphics[width=0.3\linewidth]{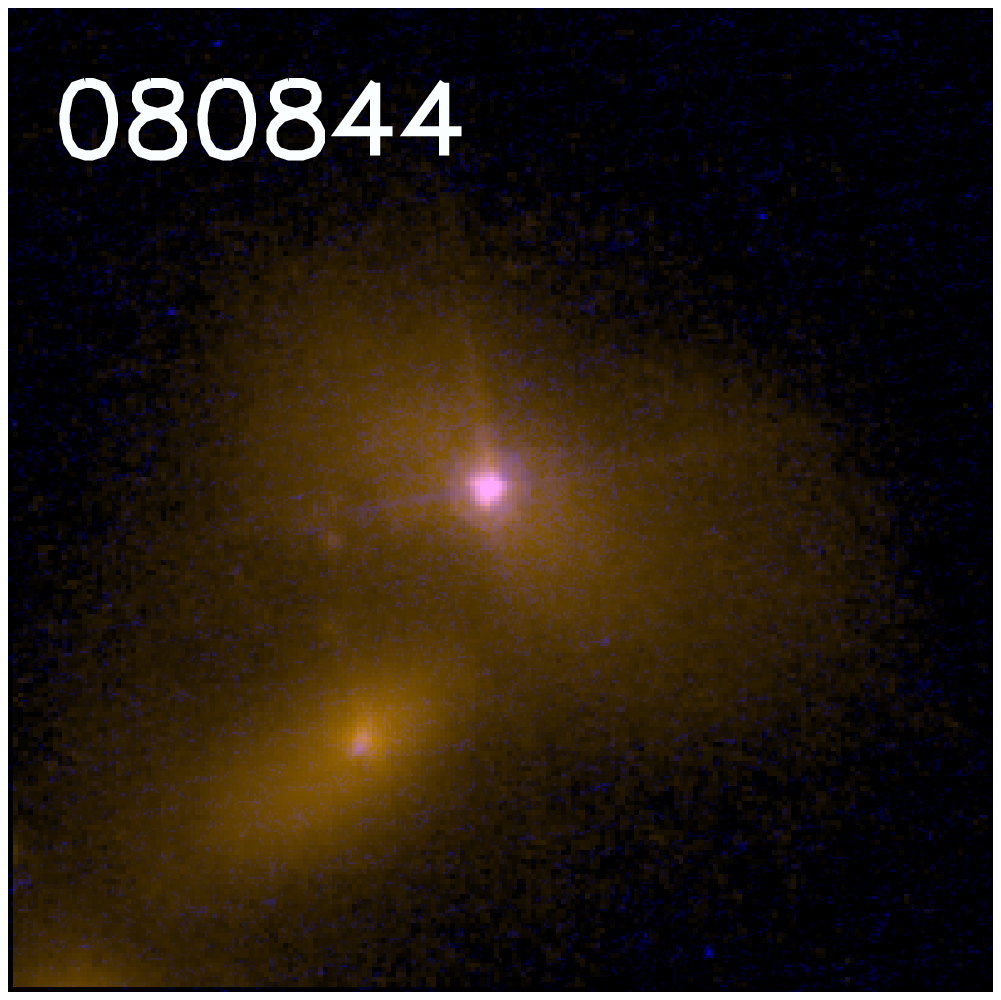}
\includegraphics[width=0.3\linewidth]{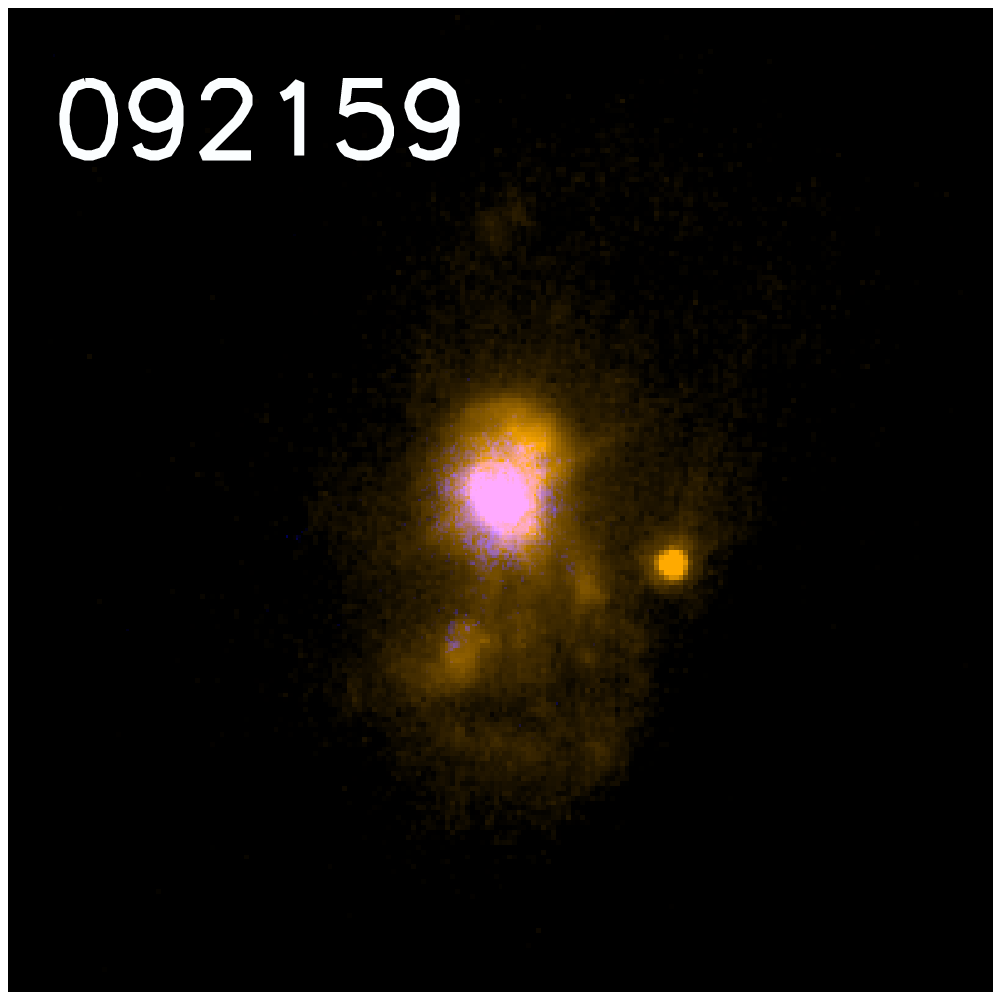}
\includegraphics[width=0.3\linewidth]{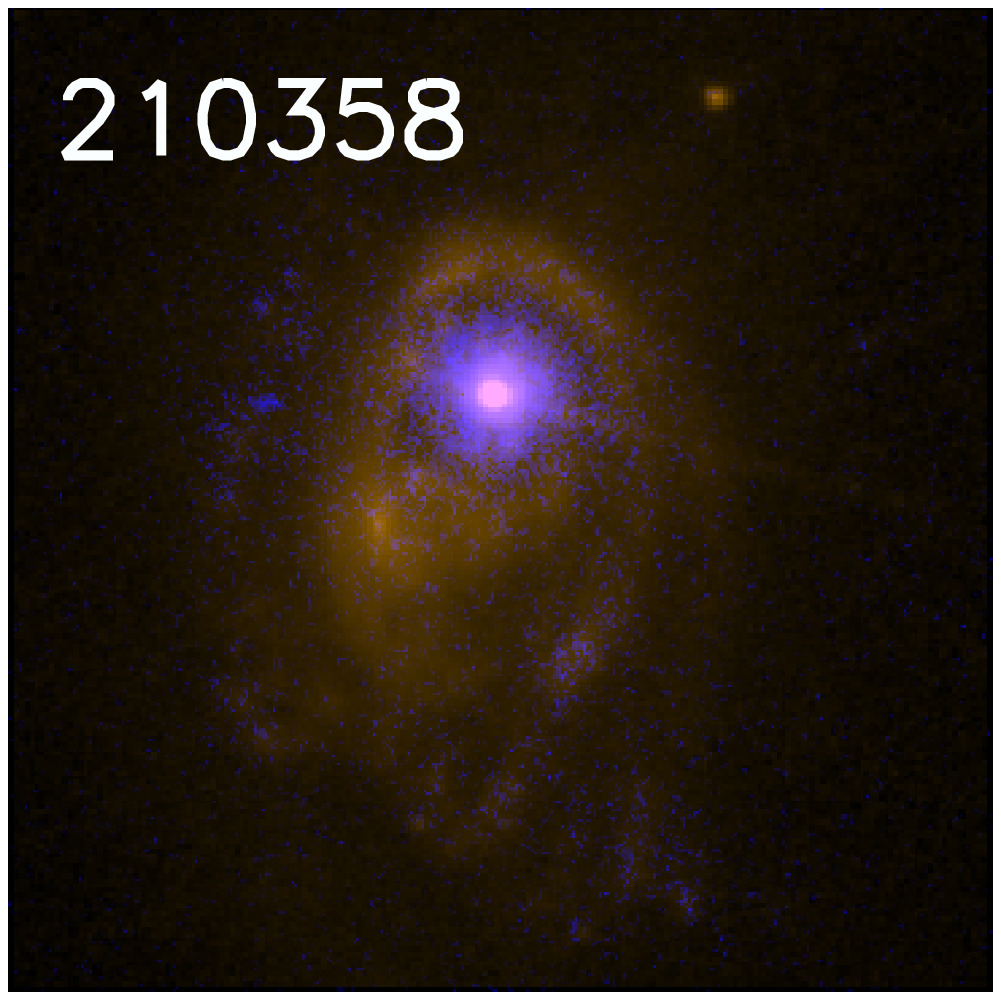}
\includegraphics[width=0.3\linewidth]{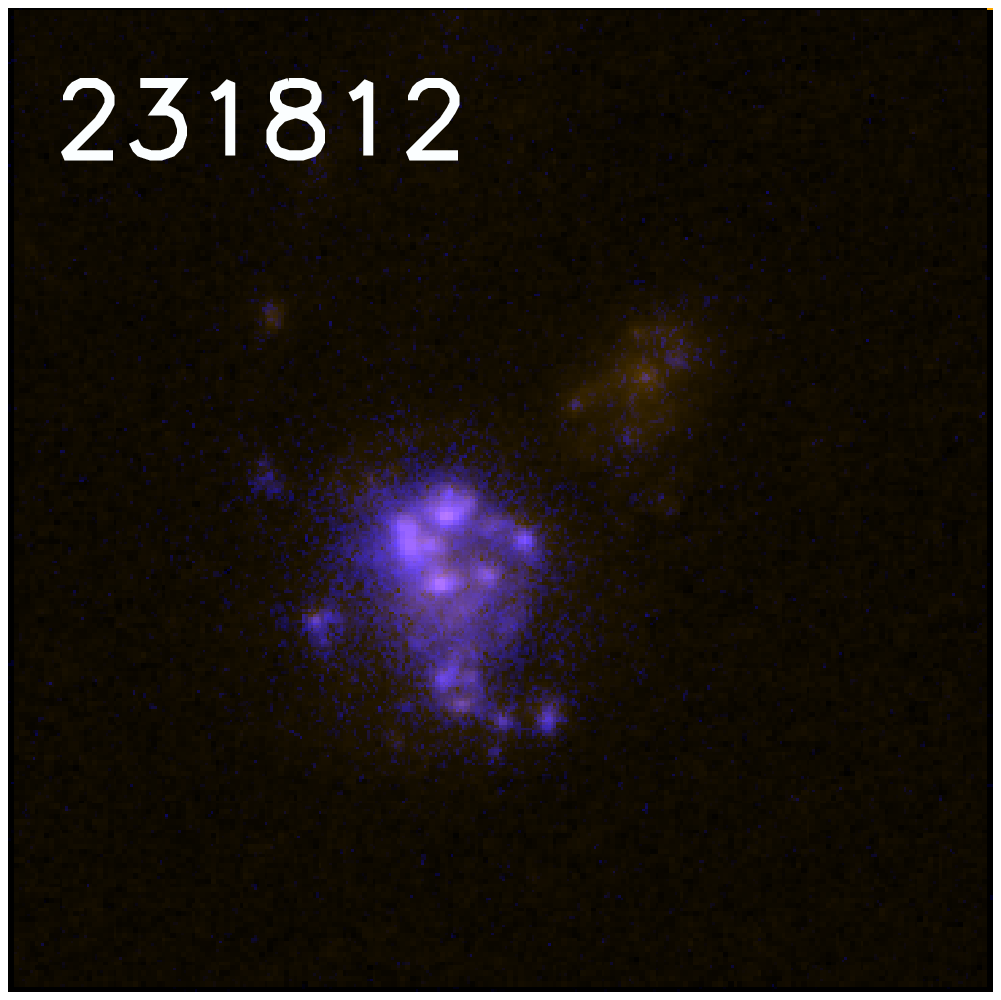}

\caption{False-colour {\it HST} images of all galaxies observed in this work. Purple represents UV emission and orange, restframe optical. Resolution is 0.1 arcsec full width at half-maximum and each image measures 6 arcsec on the side. Signatures of merging activity are evident in all objects.}\label{fig:hst}

\end{center}
\end{figure*}

Following Equation \ref{eq:gasmass}, we have inferred gas masses for all galaxies in our sample from observed SFR and optical half-light radii, and converted those values into expected CO fluxes. Table \ref{table:carma_phys_properties} summarizes the physical properties of our observed sample, and Fig. \ref{fig:hst} shows {\it Hubble Space Telescope} ({\it HST}) images of all six objects. The stellar masses were taken from the SDSS/DR7 MPA-JHU value-added catalogue\footnote{http://www.mpa-garching.mpg.de/SDSS/DR7/}. These masses were calculated by fitting a large grid of spectral synthesis models from \citet{Bruzual2003} to the SDSS u',g',r',i',z' photometry, using a \citet{Chabrier2003} initial mass function. Gas-phase metallicites were calculated using SDSS spectra from the same catalog, following the methodology in \citet{Tremonti2004}. Half-light radii are derived from {\it HST} imaging in optical bands \citep{Overzier2010}. SFR are measured from combined H$\alpha$ and MIPS 24 $\mu$m data; they typically present an uncertainty up to 0.3 dex \citep{Overzier2009}. FIR luminosities are calculated by fitting the model library of \citet{Siebenmorgen2007} to {\it Spitzer} IRAC MIPS and IRS data points \citep{Overzier2011}. Finally, dynamical time-scales are calculated by taking the ratio between half-light radii and measured circular velocities \citep{Goncalves2010}, in order to ensure consistency with similar studies at high redshift \citep{Daddi2010}.

\begin{table*}
\begin{center}
\caption{Physical properties of observed galaxies} 
\begin{tabular}{l c c c c c c c c c}\hline
ID & $z$ & log $L_{\rm FUV}$ & log $I_{\rm FUV}$ & log $L_{\rm FIR}$ & $r_{1/2}$ & SFR & log $M*$ & $12+{\rm log (O/H)}$ & $\tau_{\rm dyn}$\\
& & ($L_\odot$) & ($L_\odot$ kpc$^{-2}$) & ($L_\odot$) & (kpc) & ($M_\odot$ yr$^{-1}$)) & ($M_\odot$) & & (Myr)\\
\hline
001054 & 0.243 & 10.42 & 9.01 & ... & 4.24 & 26.9 & ... & 8.7 & 27\\
015028 & 0.147 & 10.62 & 9.38 & 11.4 & 1.83 & 50.7 & 10.3 & 8.5 & 17\\
080844 & 0.091 & 10.45 & 10.25 & 10.5 & 0.88 & 16.1 & 9.8 & 8.8 & 7\\
092159 & 0.235 & 10.82 & 9.65 & 11.6 & 1.80 & 55.1 & 10.8 & 8.8 & 13\\
210358 & 0.137 & 10.52 & 9.58 & 11.4 & 2.70 & 108.3 & 10.9 & 8.8 &15\\
231812 & 0.252 & 10.85 & 9.33 & 11.2 & 2.54 & 63.1 & 10.0 & 8.4 & 30\\
\hline
\label{table:carma_phys_properties}
\end{tabular}
\end{center}
\end{table*}

\subsection{CO observations and data analysis}\label{section:co_data_reduction}

We conducted our observations between 2010 April and 2011 June. The summary of observations, including observing date and total integration time, can be found in Table \ref{table:carma_obs}. We have selected the CO(1-0) transition ($\nu_{\rm rest} = 115.271$ GHz) as the line of choice for these observations. Although higher transitions would provide us with higher signal-to-noise ratios (S/N) for the same integration times, targeting the lowest CO transition allows us to better trace the cold molecular mass, without the need to rely on any assumptions about gas excitations levels. To complicate matters, excitation measurements for high-redshift colour-selected galaxies are scarce \citep[e.g.,][]{Dannerbauer2009,Aravena2010,Riechers2010}. At zeroth order, one could infer that CO luminosities are constant throughout all transitions, assuming the gas is thermalized. However, higher transitions require higher gas densities and may be tracing distinct regions of the ISM \citep{Harris2010}; furthermore, the gas may be subthermally excited \citep[e.g.,][]{Weiss2007,Riechers2009}. In addition, observations in the 3 mm band are simpler, since weather requirements are not as strict as observations in shorter wavelengths.

The array was used at $D$-configuration, providing a synthesized beam size of approximately 5 arcsec at the observed frequencies. This is slightly larger than the size of the observed galaxies, and we do not expect to resolve any of them at this stage of the survey. Proposed integration times were calculated aiming for a $5\sigma$ detection at the emission peak assuming the aforementioned expected CO fluxes and velocity dispersion as found in \citet{Goncalves2010}, smoothed over 10 km s$^{-1}$ channels. Actual times, however, varied according to instrument availability and weather.

\begin{table}
\begin{center}
\caption{Summary of LBAs observed by CARMA} 
\begin{tabular}{l c c c c}\hline
ID & RA & Dec & $\nu_{\rm obs}$ & Integration\\
& & & (GHz) & time (h)\\
\hline
001054 & 00:10:54.85 & 00:14:51.35 & 92.7 & 7.2\\
015028 & 01:50:28.41 & 13:08:58.40 & 100.5 & 20.7\\
080844 & 08:08:44.27 & 39:48:52.36 & 105.6 & 3.4\\
092159 & 09:21:59.39 & 45:09:12.38 & 93.3 & 5.4 \\
210358 & 21:03:58.75 & -07:28:02.45 & 101.4 & 15.3\\
231812 & 23:18:13.00 & -00:41:26.10 & 92.1 & 16.1\\
\hline
\label{table:carma_obs}
\end{tabular}
\end{center}
\end{table}

We used the MIRIAD package\footnote{The package is available at http://bima.astro.umd.edu/miriad/} to perform the CO(1--0) data analysis, choosing baseline solutions for each individual observed track. Data were mainly flagged according to variations in system temperature and resulting flux for a given gain calibrator, although specifics vary greatly on an individual basis. When data were flagged only for a fraction of a track, care was taken always to include a gain calibrator observation up to 10 min before or after each astronomical data point. In the case of multiple observed tracks for a single source, those were later combined through use of the {\it uvcat} task in MIRIAD.

Visibility files were later inverted from UV plane into real spatial/velocity flux data cubes. Velocity resolution varies according to S/N in the observations, but in most cases spectra shown here are defined in steps of 10 km s$^{-1}$, with 30 km s$^{-1}$ smoothing. `Dirty' maps were cleaned through use of the `clean' procedure, and care was taken never to use a cutoff below the observed peak flux in Jy/beam of the galaxy.

We used the CASA package to determine the total flux in each velocity channel. The integrating area was equivalent to an ellipse with same size as the beam. In Fig. \ref{fig:co_spec} we show the CO(1--0) line profile of galaxies in our sample as a function of velocity with respect to the measured systemic redshift in each case.

\section{Results}\label{sec:carma_results}



\begin{figure*}
\begin{center}
\includegraphics[width=.45\linewidth]{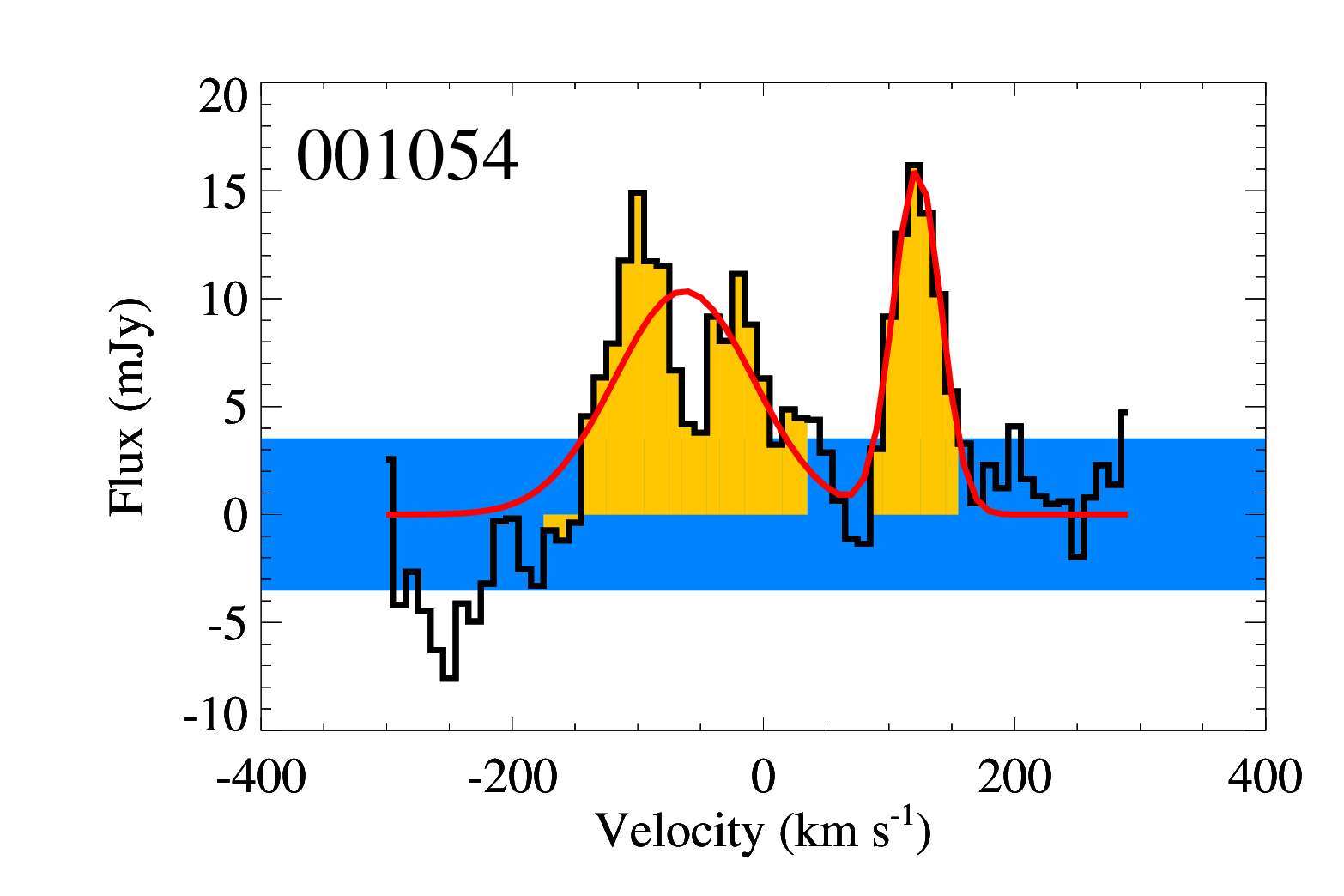}
\hskip .1 in
\includegraphics[width=.45\linewidth]{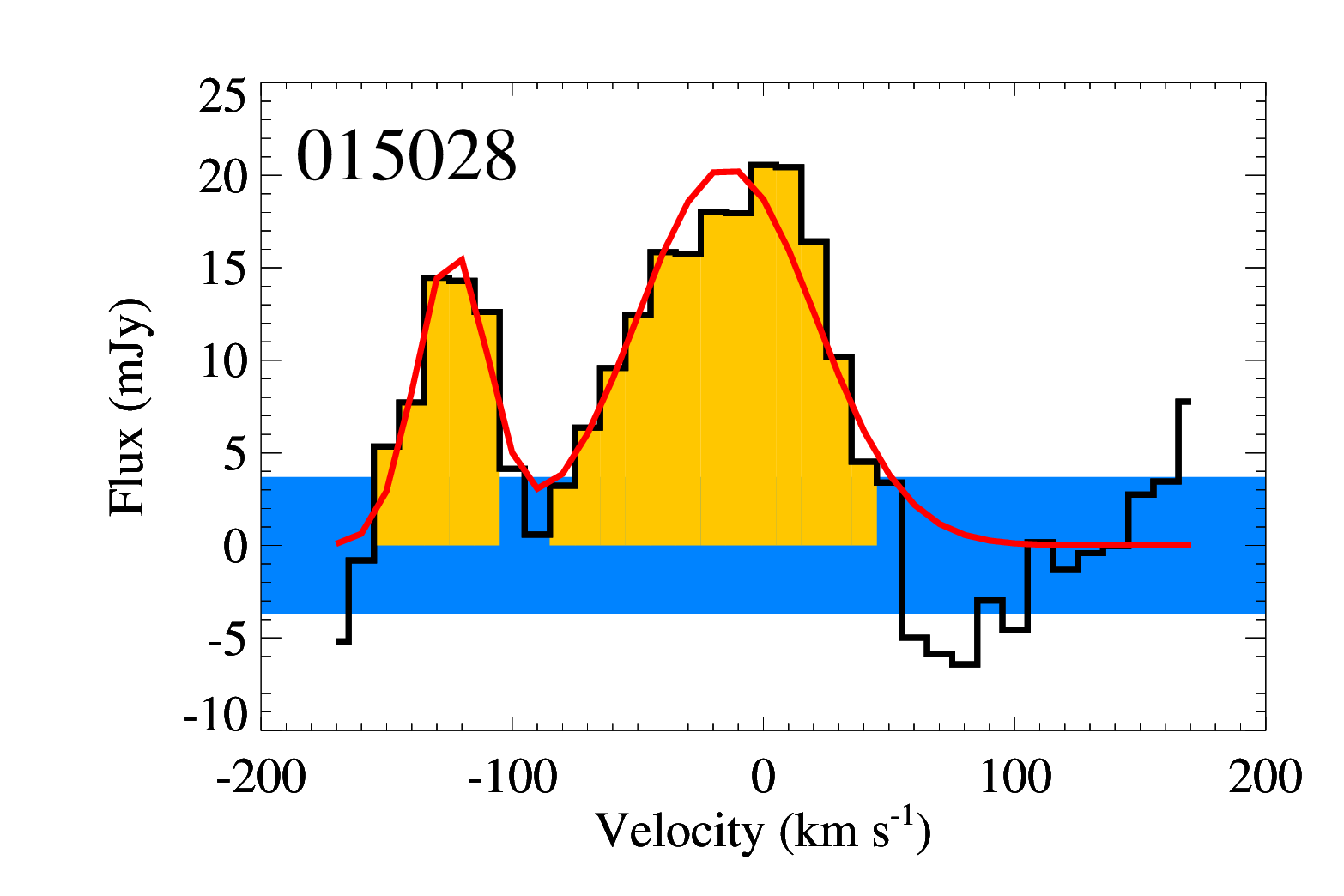}
\vskip .00 in
\includegraphics[width=.45\linewidth]{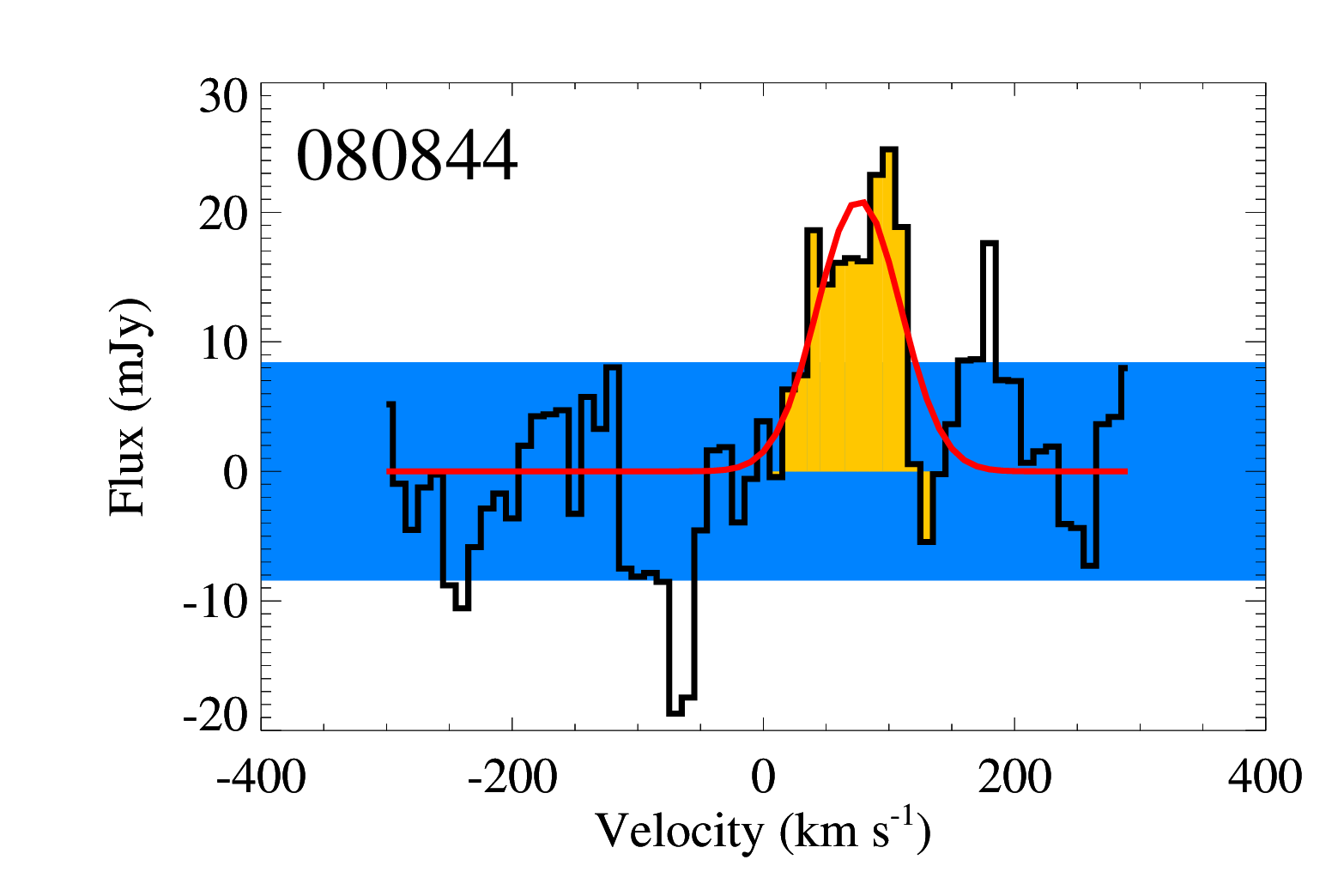}
\hskip .1 in
\includegraphics[width=.45\linewidth]{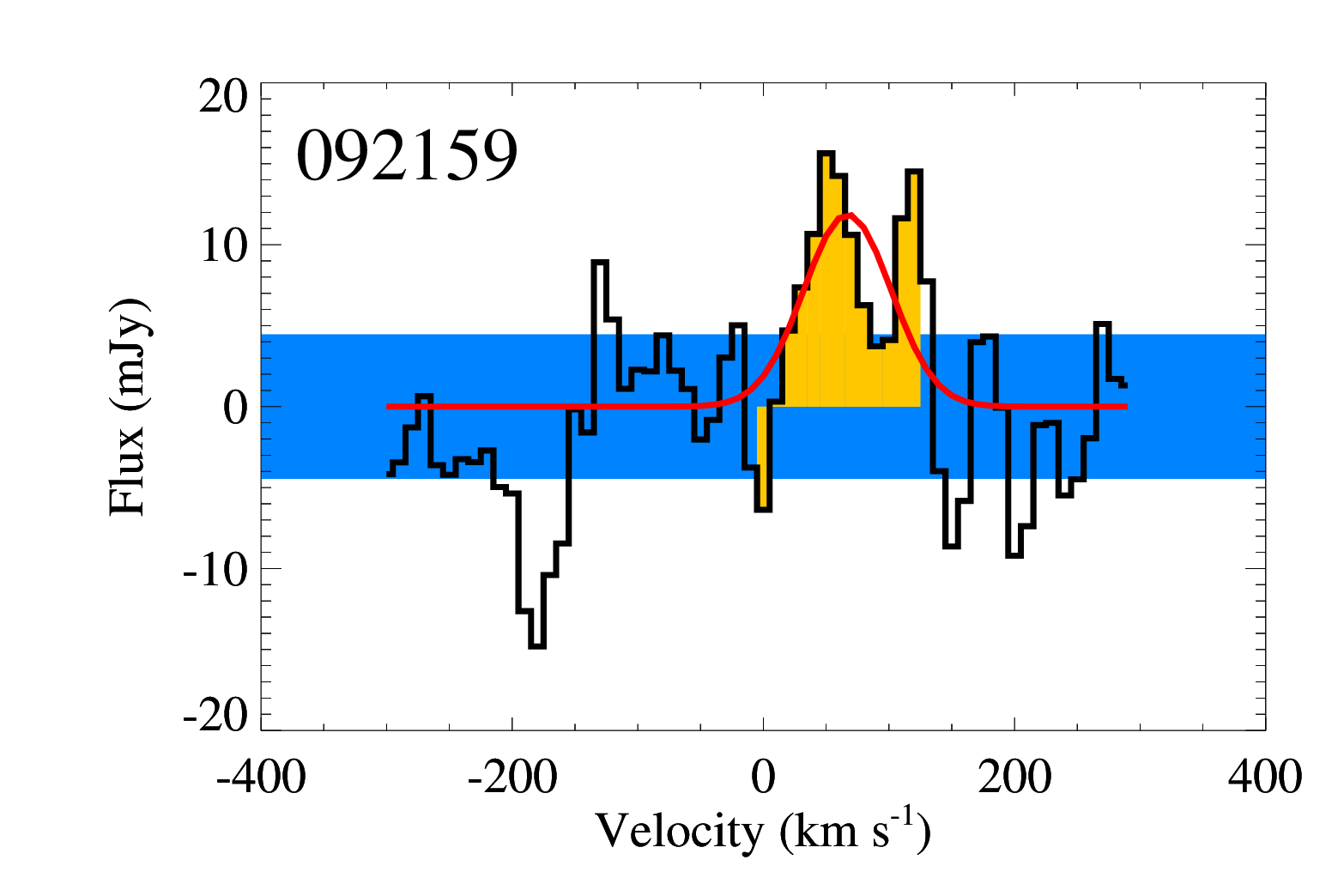}
\vskip .00 in
\includegraphics[width=.45\linewidth]{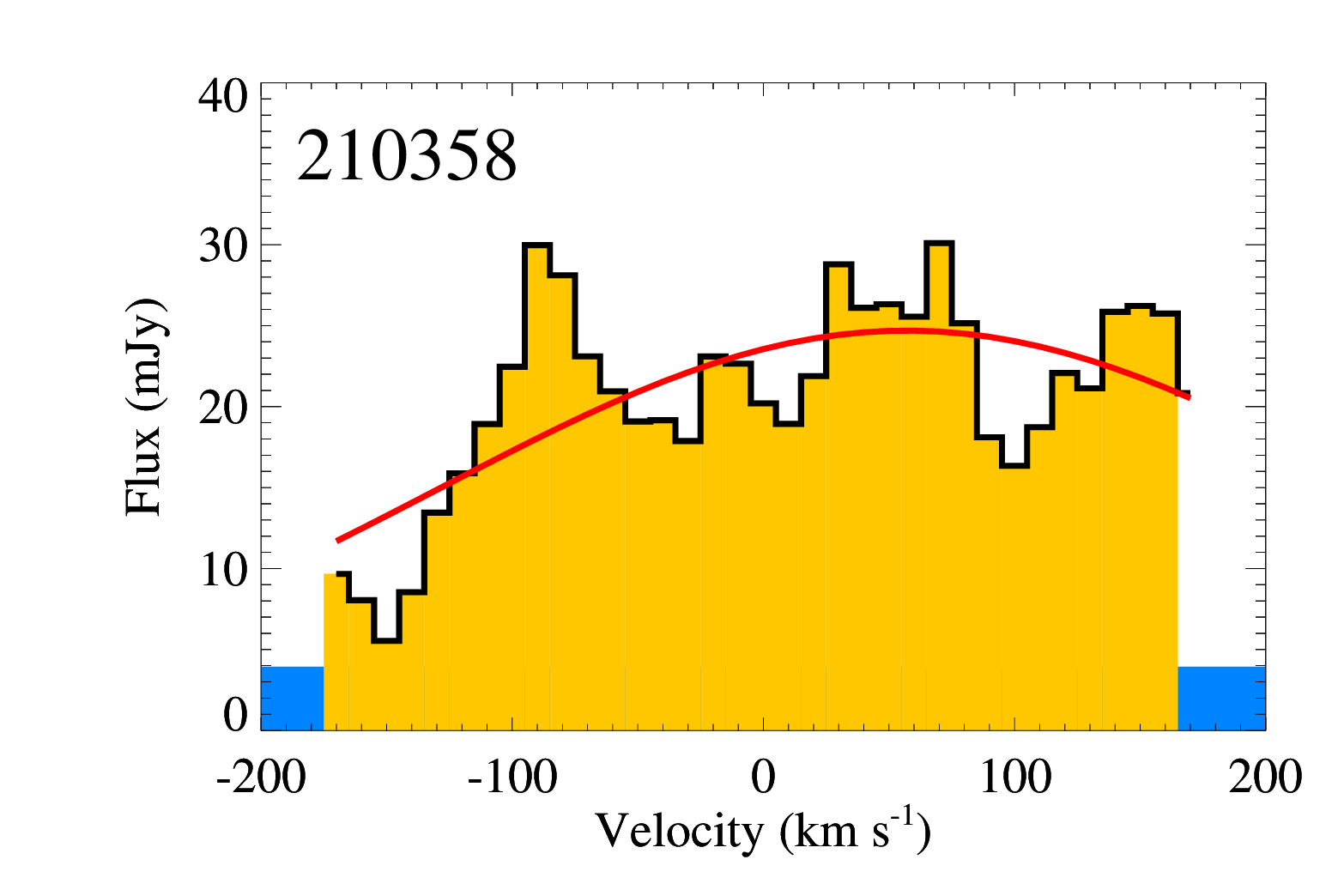}
\hskip .1 in
\includegraphics[width=.45\linewidth]{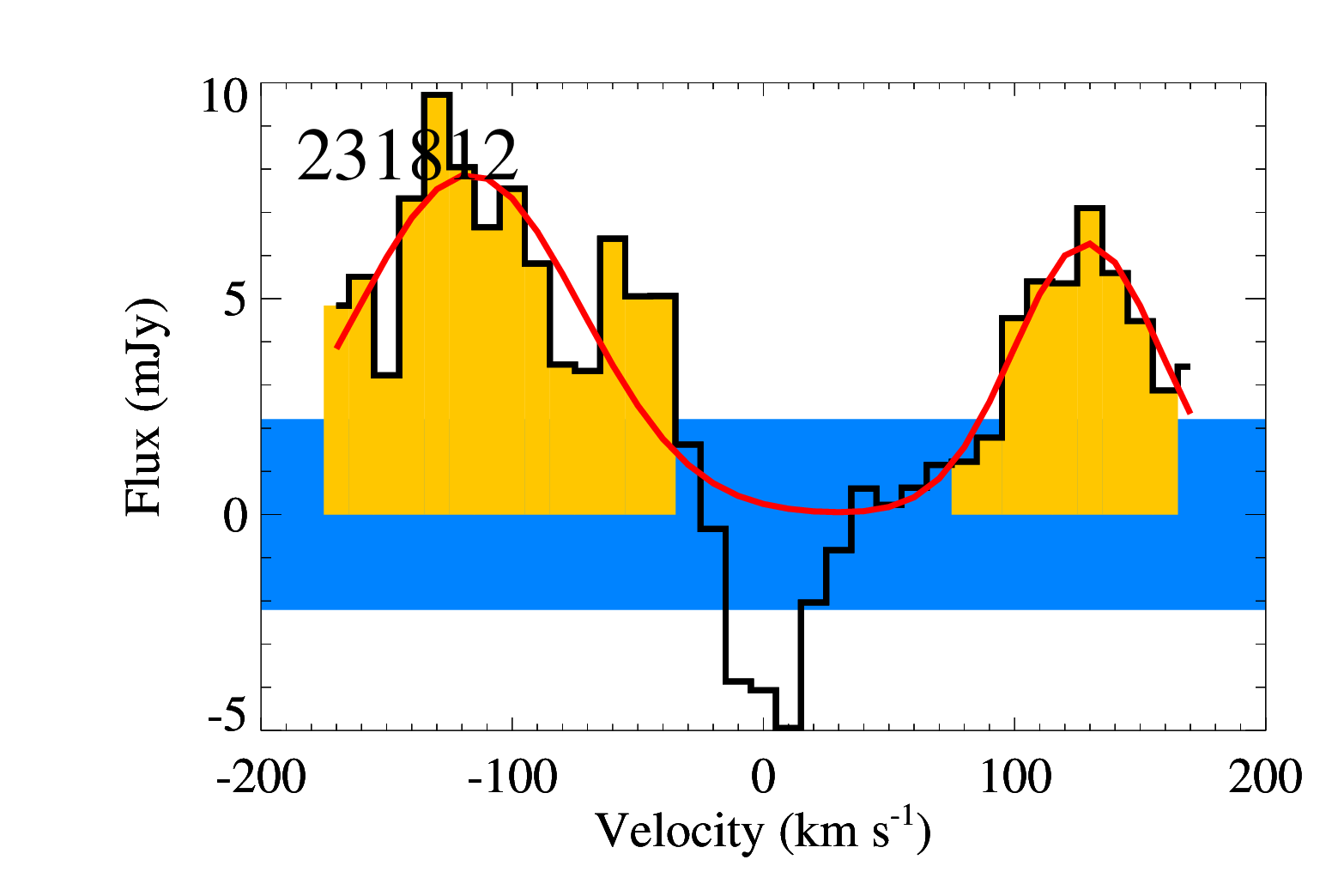}

\caption{CO(1-0) velocity profiles of detected lines. Simple or double Gaussian fits, used for flux measurements, are shown in red. Zero velocity is given by the measured spectroscopic redshift from the SDSS survey. Blue areas indicate 1=$\sigma$ dispersion values in the clean image, and we highlight in yellow spectral channels within 2$\sigma$ of the central velocity for each Gaussian.}\label{fig:co_spec}

\end{center}
\end{figure*}
\begin{figure}
\begin{center}
\includegraphics[width=\linewidth]{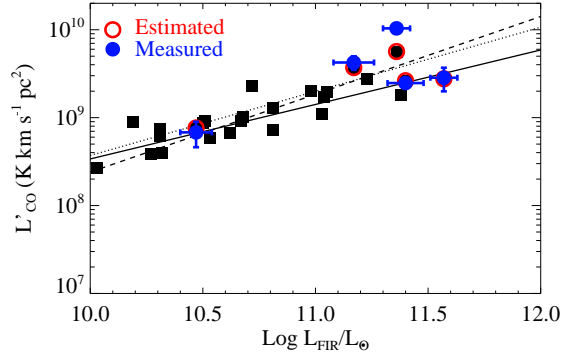}
\caption[$L'_{\rm CO}$ vs. $L_{\rm FIR}$]{LBA galaxies follow the $L'_{\rm CO}$ versus $L_{\rm FIR}$ relation described by other samples of SFGs. The solid squares represent values inferred from inverting the S-K relation, with galaxies observed with CARMA marked in red for quick comparison. Filled blue circles show actual luminosity measurements from CARMA interferometric data. The solid line is the correlation between $L'_{\rm CO}$ and $L_{\rm FIR}$ as described in \citet{Greve2005}. Also shown are relations for all high-redshift galaxies \citep[dotted line]{Carilli2013} and colour-selected galaxies at $z\sim 2$ \citep[dashed line]{Daddi2010a}.}\label{fig:lco_lfir}

\end{center}
\end{figure}

Fig. \ref{fig:co_spec} shows the spectra of all observed galaxies, in mJy. Spectral channels are binned at 10 km s$^{-1}$, and smoothed over 30 km s$^{-1}$ (three channels) each. To determine the total CO flux in the galaxy, we fit a Gaussian to the spectrum of each object. Flux is then given simply by the area below the curve. Total flux for galaxies with multiple components (such as 015028 and 231812) is given by the sum of the integrated areas of each Gaussian. All galaxies that show multiple components show clear signs of merger activity, as can be seen in Fig. \ref{fig:hst}. High-resolution interferometry should be able to decompose different spectral features into distinct spatial components. In the case of 210358, we extrapolate this integral to frequencies not covered by the observations, since bandwidth was not enough to cover the whole curve -- this might explain why the measured CO luminosity is higher than estimated (Fig. \ref{fig:lco_lfir}). We also show the spectrum for 001054, but recent work (Overzier et al, in preparation) has shown that this object is instead an AGN. We show the CO(1-0) results for this galaxy in Table \ref{table:carma_results} for completeness, but exclude it from subsequent analysis.

Three of the five remaining objects are classified as Dominant Central Objects (DCOs) by \citet{Overzier2009}, which means the presence of a very compact source in their centre: 080844, 092159, and 210358. It has been speculated that these galaxies could host AGN, since they are classified as composite starburst-AGN in typical emission line ratios used as diagnostics for nuclear activity. \citet{Jia2011} have studied the X-ray properties of DCOs and concluded that, while they do show signs of obscured AGN activity, the black hole masses estimated are low (between $10^5$ and $10^6$ M$_\odot$, compared to total masses up to $\sim 10^9$ M$_\odot$ of individual DCOs) and the bolometric luminosities are dominated by the star-forming activity. Likewise, \citet{Alexandroff2012} have observed four DCOs with the European very long baseline interferometry Network -- including all three objects in this work -- in search of point sources in the galaxy centre. Only one of them (092159) was detected, at 1.7 GHz. The black hole mass is consistent with estimates by \citet{Jia2011}, with an upper limit of $10^7$ M$_\odot$. This leads us to believe that DCOs do host the seeds of black holes commonly found in galactic bulges, but the UV light is still predominantly due to star-forming activity, generating strong superwinds with velocities up to 1500 km s$^{-1}$ \citep{Heckman2011}. In this work we treat all five objects indistinctly, but high-resolution interferometry should clarify whether the gas distribution is also concentrated in the centre of these three objects.

Fluxes are converted to CO luminosities according to the relation given by Equation \ref{eq:sco_lco}. Errors in flux measurements are the sum of the standard deviation on each velocity channel over 2$\sigma$ (where $\sigma$ is the velocity dispersion measured from the Gaussian fitting). In Table \ref{table:carma_results} we present the total CO(1--0) flux measured for each galaxy, along with inferred CO luminosities, molecular gas mass and gas fractions.

\begin{table}
\begin{center}
\caption{Results based on CARMA CO(1-0) observations}
\begin{tabular}{l c c c c}\hline
ID & $S_{\rm CO} \Delta\nu$ & $L'_{\rm CO}$ & $M({\rm H}_2)$ & $\mu_{\rm gas}$\\
& (Jy km s$^{-1}$) & ($10^9$ K km s$^{-1}$ pc$^2$) & ($10^9$ $M_\odot$) &\\
\hline
001054 & 2.19$\pm$0.41 & 6.5 & 29.7 & 0.27\\
015028 & 2.37$\pm$0.28 & 2.5 & 11.4 & 0.36\\
080844 & 1.74$\pm$0.56 & 0.7 & 3.1 & 0.33\\
092159 & 1.04$\pm$0.31 & 2.8 & 13.1 & 0.17\\
210358 & 11.5$\pm$1.5 & 10.3 & 47.7 & 0.37\\
231812 & 1.34$\pm$0.23 & 4.2 & 1.95 & 0.66\\
\hline
\label{table:carma_results}
\end{tabular}
\end{center}
\end{table}

\cite{Greve2005} have shown that there is a clear correlation between the CO luminosity of a galaxy, $L'_{\rm CO}$ and its FIR luminosity ($L_{\rm FIR}$) that can be represented by

\begin{equation}
\log L'_{\rm CO} = 0.62\log L_{\rm FIR} + 2.33. \label{eq:lco_lfir}
\end{equation}

\noindent In Fig. \ref{fig:lco_lfir} we show (as blue symbols) such correspondence between $L'_{\rm CO}$ and $L_{\rm FIR}$ for LBAs as measured with CARMA\footnote{$L_{\rm FIR}$ is defined as the luminosity integrated between 40 and 120 ${\mu}$m, and is typically a factor of 1.7 smaller than total infrared luminosity $L_{\rm IR}(3-1000\mu{\rm m})$.}. For illustrative purposes, we also show this relation for the entire LBA sample; in this case, $L'_{\rm CO}$ is inferred from a simple inversion of the S-K relation, as described by Equations \ref{eq:alpha_co} and \ref{eq:gasmass}. We notice that our sample is in very good agreement with the estimates from \cite{Greve2005}, indicating that the relation between FIR luminosities and the molecular gas reservoir in LBAs (and LBGs by extension) is not dramatically different from other galaxies of similar luminosities. Fig. \ref{fig:lco_lfir} also shows the relations between $L'_{\rm CO}$ and $L_{\rm FIR}$ determined for colour-selected galaxies as determined by \citet[dashed line]{Daddi2010a} and the combined relation for all high-redshift objects \citep[dotted line]{Carilli2013}. Given the spread in our LBA data, all models are broadly consistent with observations. This indicates that the relation between dust emission and CO luminosity is roughly the same for all galaxies, even though the LBA sample is dust deficient when compared to local SFGs with comparable bolometric luminosities \citep{Overzier2011}.

\subsection{Gas masses and fractions}

We have converted the measured CO luminosities into molecular gas masses following the relation described in Equation \ref{eq:alpha_co}. In doing so we have assumed the standard value of $\alpha_{\rm CO}=4.6$, as determined for the Milky Way galaxy. By defining the gas fraction as

\begin{equation}
\mu_{\rm gas}=\frac{M_{\rm gas}}{M_*+M_{\rm gas}},\label{eq:fgas}
\end{equation}
one can also estimate the fraction of baryonic mass in the form of gas for each galaxy. We present these values in Table \ref{table:carma_results}. We assume $M_{\rm gas}\sim M_{\rm H2}$, since most of the gas within the half-light radius will be in molecular form; we address this assumption in Section \ref{sec:atomic_molecular}.

The inferred gas depletion time-scales, given by

\begin{equation}
\tau_D\equiv M_{\rm H2}/{\rm SFR},\label{eq:tau_depletion}
\end{equation}
are between 10$^8$ and 10$^9$ yr, comparable to values typically found in high-redshift colour-selected samples \citep{Daddi2010a,Genzel2010,Tacconi2013} and lower than the canonical value of 2 Gyr inferred for galaxies in the local universe \citep{Bigiel2008,Leroy2008}. The fact that gas would be quickly depleted in these objects is often used to support the hypothesis that they are simultaneously accreting gas from the intergalactic medium, possibly through means of a ``cold-flow'' mode \citep{Dekel2009}.

\begin{figure}

\includegraphics[width=\linewidth]{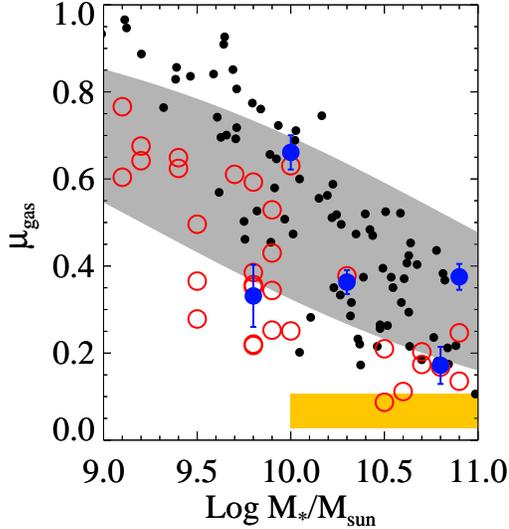}

\caption[Gas fractions as a function of stellar mass]{Gas fractions of UV-luminous SFGs are a strong function of stellar mass. Gas fractions estimated from Equations \ref{eq:gasmass} and \ref{eq:fgas}, using observed SFR, sizes and stellar masses, are shown as red hollow circles; the blue filled circles indicate actual CARMA measurements. The black points show estimates for high-redshift LBGs, from \cite{Erb2006}, calculated in the same manner. The gray shaded area represents estimated gas fractions in SFGs from the evolution of gas depletion times and sSFR between $z=1.0$ and $2.5$ \citep{Tacconi2013}, while the yellow area represents typical values measured in the local universe \citep{Saintonge2011}.}\label{fig:mstar_mu}

\end{figure}

In Fig. \ref{fig:mstar_mu} we show the LBA gas fractions as a function of stellar mass, including also additional LBAs that have not yet been observed with gas fraction values inferred from inverting the S-K relation [again following Equation \ref{eq:gasmass}, shown as red circles]. The first thing to notice is that measured values are in agreement with expectations, given the uncertainties. For comparison, we also include gas fractions estimated in the same manner for LBGs from \citet[black dots]{Erb2006}, as well as estimates at redshifts between $z=1$ and $2.5$ from the observed evolution in gas depletion times and sSFR of SFGs \citep[and references therein]{Tacconi2013}. There is a clear trend in all cases of more massive galaxies showing smaller gas fractions \citep[as found in][]{Tacconi2013}, possibly indicating these galaxies have already consumed a larger portion of their gas reservoir and converted it into stellar mass. Gas fractions are also remarkably higher than typically found in the local universe, where most SFGs have less than 10\% of their baryonic mass in the form of molecular hydrogen \citep[indicated by the yellow shaded area in Fig. \ref{fig:mstar_mu}]{Saintonge2011}. Another possibility is that star formation feedback plays an important role in regulating the infall of gas from the intergalactic medium \citep{Tacconi2013}. This trend -- especially given the high gas fractions found -- is well reproduced by our LBA sample, again reinforcing the validity of the analogy between both galaxy populations. We estimate lower gas fractions for less massive galaxies than \citet{Erb2006}; this may be due to a lower limit in detectable gas fractions in their sample.

\subsection{The Schmidt--Kennicutt relation at low and high redshift}

Whether the S-K relation holds true for high-redshift SFGs -- where higher SFR densities have been shown to exist \citep{Shapley2011} -- is still subject to some debate, as we have noted in Section \ref{sec:carma_intro}. With reliable measurement of CO(1--0) luminosities for some of the densest `normal' galaxies (i.e., not ULIRGs) ever observed, we expect that the LBA sample can shed some light on the issue.

\begin{figure}

\includegraphics[width=\linewidth]{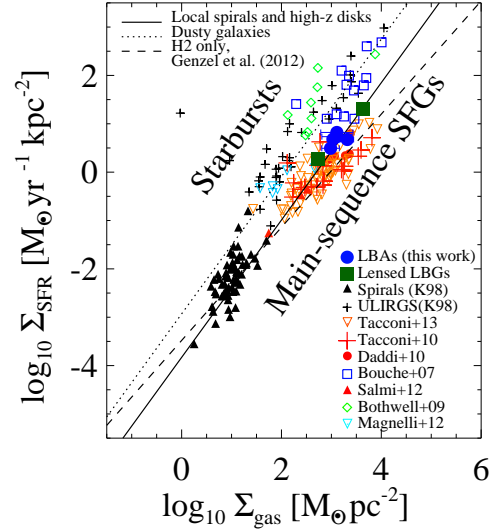}

\caption[Schmidt-Kennicutt diagram for extreme starbursts]{Star formation surface densities as a function of gas surface densities in SFGs. The two parallel lines describe the relation for regular SFGs (solid line) and ULIRG-like objects (dotted line), as described in \cite{Daddi2010}. Blue filled circles represent our observed LBA sample. Green filled squares represent the lensed LBGs studied by \citet{Baker2004} and \citet{Coppin2007}. Black crosses \citep{Kennicutt1998}, green diamonds \citep{Bothwell2009} and blue squares \citep{Bouche2007} represent ULIRGs and SMGs; black triangles \citep{Kennicutt1998}, red triangles \citep{Salmi2012}, red crosses \citep{Tacconi2010} and red circles \citep{Daddi2010} indicate spirals and disc-like galaxies at low and high redshifts. Upside-down triangles represent more recent data by \citet{Tacconi2013}; orange represents their PHIBSS data, while blue points are infrared-luminous objects originally presented in \citet{Magnelli2012}, rescaled to typical ULIRG-like $\alpha_{\rm CO}$ values. See \cite{Daddi2010}, \citet{Tacconi2013}, and references therein for details. The dashed line indicates the S-K relation for molecular gas as determined in \citet[see Section \ref{sec:atomic_molecular}]{Genzel2010}}\label{fig:daddi}

\end{figure}

In Fig. \ref{fig:daddi}, we show a compilation of the S-K relation for SFGs. The low-density end of the plot shows low-redshift spiral galaxies from the sample of \citet{Kennicutt1998}. The remainder of the points show low- and high-redshift infrared-luminous samples \citep{Kennicutt1998,Bouche2007} and `main-sequence' SFGs at $z>1$ from \citet{Tacconi2010}, \citet{Daddi2010} and \citet{Tacconi2013}. To ensure consistency, we have estimated gas surface densities for all galaxies on the lower relation (excluding submillimetre galaxies [SMGs] and ULIRGs) with a uniform $\alpha_{\rm CO}=4.6$ value, using the total gas and size measurements mentioned in the original studies. 

As we can see, the relation is apparently bimodal, as indicated by both the solid and the dashed lines. The solid line is a fit to the `main-sequence' galaxies at all redshifts, while the dashed line has the same slope, only displaced by 0.9 dex to fit through the ULIRGs and SMGs. Physically speaking, that means the latter are almost 10 times as efficient in converting cold gas into stars, although the clear separation between the relations could be a consequence of using two distinct $\alpha_{\rm CO}$ values \citep[and references therein]{Carilli2013}.

\begin{figure}

\includegraphics[width=\linewidth]{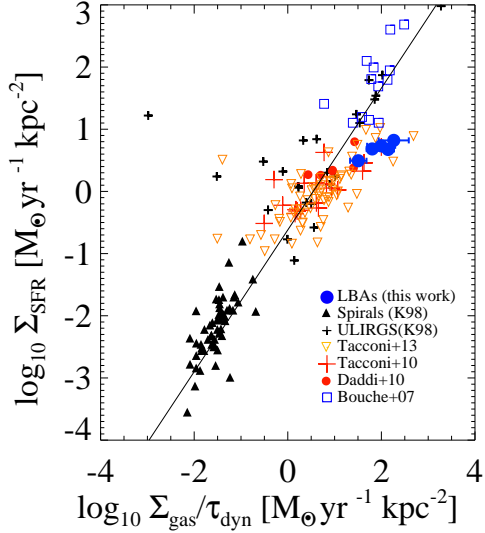}

\caption[Schmidt-Kennicutt diagram corrected for dynamical time]{Star formation surface densities as a function of gas surface densities divided by dynamical time in SFGs. Symbols are the same as in Fig. \ref{fig:daddi}, and we only include samples for which dynamical time measurements are available. The solid line is the universal relation for all SFGs as inferred by \citet{Daddi2010}. However, LBAs seem to indicate lower SFR for a given gas surface density even when dynamical times are taken into account, especially at higher densities.}\label{fig:sklaw_tdyn}

\end{figure}

In Fig. \ref{fig:sklaw_tdyn}, we show the same points, only now the abscissa values have been divided by the dynamical times in the galaxy. These time-scales are inferred from the rotational periods of the galaxies (Table \ref{table:carma_phys_properties}; for a discussion on this choice, see Section \ref{sec:dynamical}). \cite{Daddi2010a} have argued that this correction takes into account the ratio between dense gas and the more extended reservoir, since dynamical times are expected to correlate with densities as $\tau_{\rm dyn}\propto\rho^{-0.5}$ \citep{Silk1997}. In that sense, starburst galaxies have most of their gas highly concentrated in the star-forming regions, thus their star formation efficiencies are much higher. The end result is a single relation for {\it all} galaxies given by

\begin{equation}
\log {\rm SFR}/[{\rm M}_\odot \; {\rm yr}^{-1}] = 1.42\times\log({\rm M}_{H2}/\tau_{\rm dyn})/[{\rm M}_\odot \;{\rm yr}^{-1}]-0.86.
\end{equation}

Our sample, however, presents short dynamical time-scales (of order a few tens of Myr), and appear to lie below the relation inferred by \citet{Daddi2010}. This might be a result of how dynamical time-scales are defined; we discuss in Section \ref{sec:dynamical} whether the choice of the rotational periods of galaxies as dynamical time-scales is valid for all samples.

\section{Discussion}\label{sec:carma_discussion}

One must be careful to take into account possible caveats in order not to over-interpret the data. In this section we list possible caveats that might affect how we view our results: the distinction between atomic and molecular gas, the adopted definition for dynamical time-scale and the $\alpha_{\rm CO}$ conversion factor.

\subsection{Atomic and molecular hydrogen}\label{sec:atomic_molecular}

First of all, one needs to be careful regarding total gas versus the molecular gas alone. At high redshifts, in particular, where the direct detection of atomic hydrogen is nearly impossible, it seems that {\it total} and {\it molecular} gas are used indistinguishably.

In fact, at the observed scales, it is true we expect the molecular gas to dominate the baryonic mass in the galaxy. As \citet{Bigiel2008} pointed out, the surface density of HI saturates at about 10 M$_\odot$ pc$^{-2}$, at least in the local universe. In our LBA sample the surface density of (molecular) gas is about a hundred times higher; therefore we would expect the atomic hydrogen to dominate the bulk of the mass only at distances greater than 10 times its optical radii, assuming constant surface density at this saturation level over the whole area.

Still, one needs to be careful to distinguish between both definitions. While \citet{Daddi2010} have made comparisons with the S-K relation for {\it total} gas, \citet{Genzel2010}, for example, determine the relation between star formation surface density and the surface density of {\it molecular} gas, finding an exponent closer to the linear ($N=1$) relation found for molecular gas only by \citet{Bigiel2008}, the reason being that the molecular component alone is responsible for star formation:

\begin{equation}
\log {\rm SFR}/[{\rm M}_\odot \; {\rm yr}^{-1}] = 1.17\times\log({\rm M}_{H2})/[{\rm M}_\odot] - 3.48.
\end{equation}
This is indicated as the dashed line in Fig. \ref{fig:daddi}. While this latter relation seems more appropriate for the galaxies described in \citet[yellow crosses]{Tacconi2010}, the S-K relation for LBAs is more accurately described by Equation \ref{eq:sklaw}. Whether this distinction is physical -- since our study includes less massive, more UV-luminous galaxies -- or purely a systematic effect due to sample selection is yet unclear. A survey of identical CO transitions [the (3--2) line in most high-redshift studies, instead of (1--0)] could help clarify this difference, as well as a study of a larger sample of UV-luminous SFGs.

\subsection{The dynamical time-scale}\label{sec:dynamical}

Likewise, we need to ensure that we understand the meaning of dynamical time in these galaxies. In principle, the dynamical time represents the amount of time a single star takes to orbit around the galaxy. This time-scale can be linked to the duration of processes that can spread throughout a given radius; for instance, a simple model for star formation in a galaxy through the collapse of cold molecular gas requires the cooling time to be shorter than the dynamical time-scale within a given radius.

In the case of extreme SFGs, however, the definition of the dynamical time-scale is not so clear. In Fig. \ref{fig:sklaw_tdyn}, for example, all `main-sequence' galaxies at high redshift (and our own LBA sample, for consistency) have dynamical time-scales defined as the rotation time-scale at the half-light radius, while for the local sample of spirals and ULIRGs \citep{Kennicutt1998} this quantity is defined at the outer radii of galaxies.

More importantly, though, we question the validity of characterizing dynamical time-scales in all SFGs by use of circular velocities. Velocity dispersions in SMGs, in particular, are remarkably high, so perhaps using $t_{\rm dyn}\sim r/\sigma$ might be a better description \citep[for a detailed discussion of the resolved gas kinematics in SMGs, see][]{Menendez-Delmestre2013}. In the case of LBAs, \citet{Goncalves2010} have shown that not all galaxies can be accurately described as rotating discs. Likewise, \citet{ForsterSchreiber2009} have shown that about 1/3 of their SINS sample are dispersion dominated. Since most of our galaxies present $v/\sigma$ values smaller than 1, a different definition of $t_{\rm dyn}$ would place our data on the same relation as \citet{Daddi2010}. Furthermore, it has been found that there is a correlation between stellar masses and the amount of rotational dynamical support \citep{Goncalves2010,NewmanS2013}; this result, combined with the steepness of the stellar mass function at high redshift \citep{Reddy2009}, would indicate that the percentage of dispersion-dominated galaxies in the high-redshift population could be even higher.

According to \citet{Krumholz2012}, the time-scales for high-redshift galaxies is related to the free-fall time for gas, which is dominated not by internal dynamics of giant molecular clouds, but instead by the bulk motion of the ISM. While the formation of instabilities in a rotating disc can describe well the advent of clumps in massive galaxies, we argue that turbulent motion makes for a more complex scenario in smaller irregular objects, in particular given strong indications that turbulent pressure might be responsible for self-regulating star-formation in these cases \citep{LeTiran2011,Lehnert2013}.

\subsection{The CO$\rightarrow$ H$_2$ conversion factor}\label{section:co_to_h2}

One of the key ingredients in determining the total molecular gas mass in galaxies is converting CO luminosities into molecular hydrogen masses. Although a typical value of $\alpha_{\rm CO}=4.6$ M$_\odot$ $($K km s$^{-1}$ pc$^2)^{-1}$ appears appropriate for normal SFGs in the local universe, the picture is less clear regarding low-metallicity objects and compact SFGs at high redshift.

On one hand, \cite{Leroy2011} have found that, in the Local Group, low-metallicity galaxies such as the Small Magellanic Cloud show $\alpha_{\rm CO}$ values as high as 70 -- at this level, CO is no longer self-shielded from ionizing radiation in the galaxy, while hydrogen is still present in molecular form; therefore CO emission is much fainter for the same amount of gas \citep{Wolfire2010}. On the other hand, dusty starburst galaxies typically show $\alpha_{\rm CO}$ values closer to $0.8$, probably due to turbulence in the ISM \citep{Downes1998, Papadopoulos2012}. For a thorough review on the subject, we refer the reader to \citet{Bolatto2013}.

This presents a particular conundrum for the study of molecular gas in LBAs; while in the local universe low-metallicity galaxies are typically dwarf irregulars and compact starbursts tend to be (dusty) LIRGs or ULIRGs, many LBAs can be classified {\it both} as low-metallicity and compact, dense SFGs. On one hand, as discussed in Section \ref{sec:carma_sample}, many LBAs -- and 4 out of the 5 objects presented here, see Fig. \ref{fig:lco_lfir} -- are LIRGs, with IR luminosities above L$_{\rm FIR} > 10^{11}$ L$_\odot$, and all of them would be considered as `starburst' galaxies in the local universe. \citet{Narayanan2012} have shown that increased gas temperatures and high velocity dispersions, both of which are clearly present in LBAs, could have the effect of decreasing the $\alpha_{\rm CO}$ factor to values more similar to infrared-luminous merger-dominated starbursts. On the other hand, LBAs present lower metallicities and extinction values than galaxies within the same stellar mass range, and that would drive $\alpha_{\rm CO}$ values up. We emphasize that this combination of physical conditions of the ISM should be prevalent in SFGs in earlier epochs. 

This determination is particularly important because the bimodality observed in the S-K relation can be attributed, at least in part, to the use of different $\alpha_{\rm CO}$ values for dusty and UV-luminous samples.  In the case of high-redshift galaxies, direct measurements of the conversion factor are nearly impossible, and one has to rely on empirical results for low-redshift galaxies, which present dramatically different ISM properties and might misrepresent the problem.

As an example, \cite{Genzel2012} have attempted to estimate $\alpha_{\rm CO}$ in high-redshift galaxies, assuming gas masses from simple inversion of the S-K law (e.g. Equation \ref{eq:gasmass}) and comparing with observed CO luminosities. The authors then infer a power-law relation between metallicity and $\alpha_{\rm CO}$, by using measured values from \cite{Leroy2011}. The resulting empirical relation is as follows:

\begin{equation}
\log\alpha_{\rm CO} = -1.3 \times \mu_0 + 12.1, \label{eq:alpha_genzel}
\end{equation}
where $\mu_0$ is the metallicity as defined by \citet{Denicolo2002}. However, whether that parallel between local spirals and high redshift galaxies is valid is debatable. We know that at $z\sim2$, densities, metallicities and stellar masses, amongst other observables, are distinct from those seen in local galaxies, and we might expect that $\alpha_{\rm CO}$ for a low-metallicity galaxy will be smaller at $z=2$ then at $z=0$ due to a denser and more turbulent ISM, as discussed above.

\begin{figure}

\begin{center}
\includegraphics[width=\linewidth]{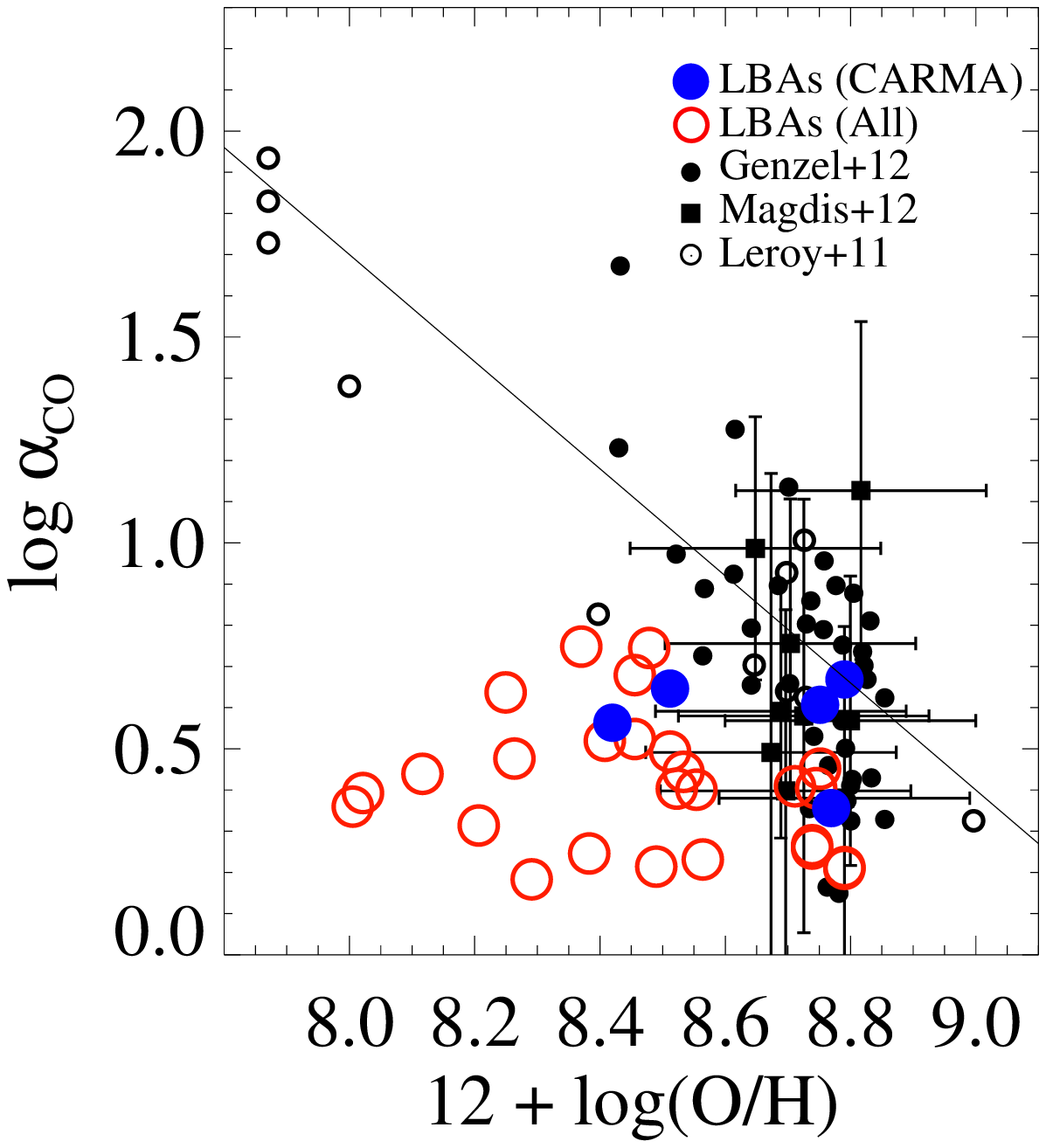}

\caption[$\alpha_{\rm CO}$ as a function of metallicity]{$\alpha_{\rm CO}$ as a function of metallicity. Coloured symbols indicate our LBA sample; red circles represent the observed objects, while for the rest of our sample (blue circles) we infer CO luminosities from $L_{\rm FIR}$. Filled black symbols are the high-redshift objects from \citet[circles]{Genzel2012} and \citet[squares]{Magdis2012}. Hollow small circles are the local galaxies from \citet{Leroy2011}. All metallicity measurements are reduced to the same methodology of \citet{Denicolo2002}. The solid line represents the empirical relation found in \citet{Genzel2012} using all samples; low-metallicity LBAs appear to lie below that relation by a factor of $\sim\;$5.}\label{fig:alpha_lbas}

\end{center}
\end{figure}

To illustrate this statement, we compared estimated values for $\alpha_{\rm CO}$ in all objects in our sample [estimating gas masses from Equation \ref{eq:gasmass} and CO luminosities from Equation \ref{eq:lco_lfir}; Fig. \ref{fig:alpha_lbas}]. We converted between metallicities following the empirical relation found in \citet{Kewley2008}. We then compared these estimates with values found in \citet{Genzel2012} and \citet{Magdis2012}. At high metallicities there is good agreement between our sample and the high-redshift one; however, at low metallicities, our inferred $\alpha_{\rm CO}$ are smaller than local galaxies with similar metallicities almost by a factor of 5. For galaxies observed thus far with CARMA (red points), most of which are more metal rich, we conclude that the Galactic value for $\alpha_{\rm CO}$ is appropriate. The estimates for lower metallicity galaxies still carry considerable uncertainties, given that we have not directly measured CO luminosities in these objects yet.

\begin{figure}

\begin{center}
\includegraphics[width=\linewidth]{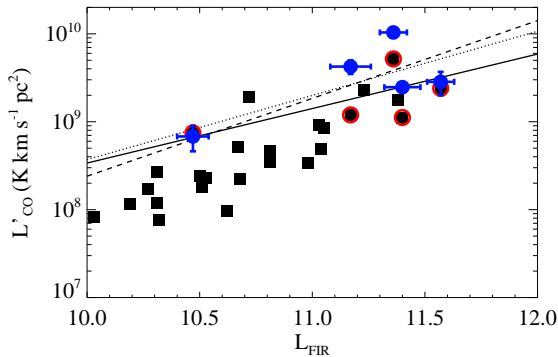}

\caption[$L'_{\rm CO}$ vs. $L_{\rm FIR}$ in LBAs with metallicity-dependent $\alpha_{\rm CO}$]{Same as Fig. \ref{fig:lco_lfir}, but following the metallicity dependence of $\alpha_{\rm CO}$ described in \cite{Genzel2012}.}\label{fig:lco_lfir_alpha}

\end{center}
\end{figure}

Another way to look at the problem is pictured in Fig. \ref{fig:lco_lfir_alpha}, where we show values of $L'_{\rm CO}$ versus $L_{\rm FIR}$, as in Fig. \ref{fig:lco_lfir}, but this time we infer CO luminosities assuming an $\alpha_{\rm CO}$ conversion factor following Equation \ref{eq:alpha_genzel}. We notice that the inferred CO luminosities are much lower than expected for a given FIR luminosity, because at lower luminosities (i.e., lower metallicities) the same gas masses would produce much fainter CO emission. The conclusion is that one of three assumptions must be wrong: either (1) the $L'_{\rm CO} - L_{\rm FIR}$ is not valid for LBAs; (2) the S-K relation is not valid for LBAs; or (3) the $\alpha_{\rm CO} - Z$ relation is not valid for LBAs. The {\it Atacama Large Millimetre/Submillimetre Array} (ALMA) should be able to provide a definite answer to this question, since measurement of CO(1--0) with the finished array is achievable after less than an hour of integration even for the least massive, most metal-poor objects in our sample. The study of low-mass dust-deficient galaxies is of particular relevance, since these objects are believed to dominate the star formation density of the universe at high redshifts \citep{Reddy2009}.

\section{Summary}\label{sec:carma_summary}

We have initiated a survey measuring CO(1--0) emission in Lyman break analogues, in an attempt to estimate gas fractions in these galaxies and obtain further insight towards the relation between star formation and the cold gas reservoir in UV-luminous SFGs. We discuss expectations using well-established relations for the local universe (such as the S-K relation) and observed surface densities and gas fractions of high-redshift SFGs.

Using the CARMA interferometer, we have detected five objects at high S/N. We find CO luminosities that agree well with expectations from $L_{\rm FIR}$ (Fig. \ref{fig:lco_lfir}). These galaxies show strong emission, indicative of high gas masses and gas fractions, between 20 and 60\% (Fig. \ref{fig:mstar_mu}). They also show shorter gas consumption time-scales than typically found in SFGs in the local universe (between 0.1 and 1 Gyr). These values agree with our expectations, further supporting our use of LBAs as low-redshift proxies for exploring the connections between star formation, molecular gas and triggering mechanism in sources that are more similar to typical high redshift galaxies than any other local samples studied thus far.

Furthermore, we have shown that LBAs follow the local S-K relation, albeit at much higher surface densities than typical spirals at $z\sim 0$ (Fig. \ref{fig:daddi}). This is in accordance with other high-redshift populations \citep{Daddi2010a}, and the result distinguishes our sample from infrared-luminous objects such as ULIRGs and SMGs. The authors in that work have argued that the bimodality can be removed if one takes into account the dynamical times of each object. This is still uncertain in the case of our galaxies; there is an apparent shift to smaller SFR surface densities even when taking into account dynamical times, but this may be the result of the use of rotational periods as the dynamical time-scales of galaxies.

We note the inherent difficulties in determining gas masses from CO luminosities, in particular given the uncertainties in $\alpha_{\rm CO}$ (especially at low metallicities). Given that low- and intermediate-mass galaxies will be more metal poor than similar objects in the local universe, this could strongly affect our understanding of molecular gas at earlier epochs. Our LBA sample, on the other hand, offers an excellent opportunity to study molecular gas in low metallicity, compact SFGs and ALMA observations will allow for accurate estimates of $\alpha_{\rm CO}$ in such objects.

\vskip .5 in

We thank the anonymous referee for suggestions that helped improve this paper. We would also like to thank Andrew Baker and Tim Heckman for useful comments. TSG gratefully acknowledges CAPES for financial support. Support for CARMA construction was derived from the states of California, Illinois, and Maryland, the James S. McDonnell Foundation, the Gordon and Betty Moore Foundation, the Kenneth T. and Eileen L. Norris Foundation, the University of Chicago, the Associates of the California Institute of Technology, and the National Science Foundation. Ongoing CARMA development and operations are supported by the National Science Foundation under a cooperative agreement, and by the CARMA partner universities.

\bibliography{CO}

\label{lastpage}

\end{document}